\documentclass[aps,pra,superscriptaddress,amsmath,amssymb,showpacs]{revtex4-2}
\usepackage{amssymb,bm}
\usepackage{graphicx}
\usepackage{amsmath}
\usepackage{color}
\usepackage[colorlinks=true,citecolor=blue,urlcolor=blue]{hyperref}
\begin{document}

\title{Quadrupole radiation and $e^+e^-$ pair production  in the collision of nonrelativistic nuclei.}

\author{I.V. Obraztsov}
\affiliation{Budker Institute of Nuclear Physics of SB RAS, 630090 Novosibirsk, Russia}
\affiliation{Novosibirsk State University, 630090 Novosibirsk, Russia}
\author{A.I.Milstein}
\affiliation{Budker Institute of Nuclear Physics of SB RAS, 630090 Novosibirsk, Russia}
\affiliation{Novosibirsk State University, 630090 Novosibirsk, Russia}

\date{\today}

\begin{abstract}
We analyze  the one-photon mechanism of $e^+e^-$  pair production in the collision of nonrelativistic nuclei. The contribution of electric quadrupole  radiation of virtual photon to the corresponding cross section is found. The effect of the finite nuclear size  is considered in detail. A comparison is made with the contribution of electric dipole radiation of virtual photon and with the contribution of the two-photon pair production mechanism. It is shown that the   contribution of electric quadrupole radiation is dominant in a wide range of relative velocities. The cross section for the production of  $e^+e^-$ pair with the capture of an electron by one of the nuclei is also analyzed.	

\end{abstract}

\maketitle

\section{Introduction}
The study of the process of $e^{+}e^{-}$ pair production in  nuclear collision began in the 1930s, since this process is important and interesting both from the theoretical and experimental point of view. The qualitative picture of the process differs significantly at high and low energies of colliding nuclei.
At high energies, the pair production cross section was considered in Refs.~\cite{LandLif1934,Racah1936} in the leading approximation in the parameters $\eta_{1,2}=Z_{1,2}\,\alpha$. Here $Z_{1,2}$  are the charge numbers of nuclei, $\alpha$ is the fine-structure constant, the system of units $\hbar=c=1$ is used. In the 1990s, numerous works were devoted to the discussion of the Coulomb corrections to the pair production cross section at high energies (the contribution of higher orders of perturbation theory in the parameters $\eta_{1,2})$, see \cite{BaltMcL1998,IvaScSe1999,EiReScG1999,SegeWel1999,Lee2000} and reviews \cite{BauHeTr2007, UFNKM2019}. At high energies, the interaction between nuclei can be neglected, so that the nuclei can be considered as moving with certain impact parameters at a constant velocity. Each nucleus emits a virtual photon, and $e^{+}e^{-}$  pair appears as a result of the interaction of these photons. We will conventionally call this mechanism a two-photon pair production mechanism. The corresponding cross section of the process depends on the relative velocity $\beta$ between the nuclei and does not depend on their masses $M_{1,2}$. In the cited works, the cross sections of the processes in which the electron and positron are in the states of a continuous spectrum (free-free pair production)  and in which the electron is captured by one of the nuclei (bound-free pair production) were analyzed. A feature of the pair production process at high energies is that the characteristic impact parameters $\rho$ of colliding nuclei are large (or comparable) with the Compton electron wavelength  $\lambda_C=1/m_e$, where $m_e$ is the electron mass.

At  small relative velocity $\beta\ll 1$ between nuclei, the Sommerfeld parameter $\eta_0=Z_1Z_2\alpha/\beta=\eta_1\eta_2/(\beta\alpha)$ becomes large already for relatively light nuclei, and the interaction between nuclei cannot be neglected. At $\eta_0\gg 1$, the interaction can be taken into account in the quasiclassical approximation. In addition, the characteristic impact parameter  $\rho$, which makes the main contribution to the pair production cross section, becomes much smaller than $\lambda_C$, so that   it is necessary to take into account the finite sizes $R_{1,2}$ of the  nuclei. At small $\beta$, the one-photon pair production mechanism becomes important. This mechanism corresponds to radiation of a virtual photon due to scattering of nuclei with subsequent conversion of this photon into $e^{+}e^{-}$ pair. The contribution of the one-photon mechanism to the cross section  of free-free pair production was studied in \cite{Lif1935, Okun1955, Greiner1976}, see also \S 100 in \cite {BLP}. 

Numerous papers are devoted to the bound-free $e^+e^-$  pair production  in the collision of slow heavy nuclei ($ Z_1 + Z_2> 173 $). Interest in this issue is related to the discussion of the spontaneous $e^+e^-$ pair production in superstrong electric fields; the corresponding references can be found in the review \cite{Greiner1985}  and in the recent work \cite{shabaev2020}.

In Ref.~\cite{LeeMingulov2016}, the contribution of the two-photon mechanism to the total cross section of pair production was obtained in the lowest Born approximation for arbitrary $\beta $ without taking into account the interaction between nuclei. It turned out that for $\beta\ll 1$ this contribution is strongly suppressed by the factor $\beta^8$. As a result, the question arose about the magnitude of the Coulomb corrections due to the interaction of  produced pair with the nuclei. This problem was solved in Ref.~\cite{LM2016} in which the two-photon contributions to the free-free pair production cross section $\sigma_{ff}^{\gamma\gamma}$  and to the bound-free  pair production cross section $\sigma_{bf}^{\gamma\gamma}$ were obtained at $\beta\ll 1$ and $\eta_{1,2}\ll 1$ with account for the Coulomb corrections. These results read
\begin{align}
&\sigma_{ff}^{\gamma\gamma}=\frac{(\eta_{1}\eta_{2})^{2}\beta^{6}}{1050\pi m_e^{2}}\Bigg\{\pi^{2}\left(\eta_{1}+\eta_{2}\right)^{2}+\frac{592}{105}\beta^{2}\Bigg\}\,,\nonumber\\
&\sigma_{bf}^{\gamma\gamma}  =\frac{16(\eta_{1}\eta_{2})^{2}\left(\eta_{1}^{3}+\eta_{2}^{3}\right)\beta^{6}}{15015m_e^{2}}\zeta_{3}\left\{ \pi^{2}\left(\eta_{1}+\eta_{2}\right)^{2}+\frac{976}{153}\beta^{2}\right\} \,,
\end{align}
where  $\zeta_{3}=\sum_{n=1}^{\infty}\frac{1}{n^{3}}\approx 1.2\,$.
It is seen that for $1\gg\eta_1+\eta_2\gg \beta$ the Coulomb corrections exceed the Born contribution and are suppressed as $\beta^6$, and not as $\beta^8$. In addition,  $\sigma_{ff}^{\gamma\gamma}\gg\sigma_{bf}^{\gamma\gamma} $  for $\eta_{1,2}\ll 1$. Note that for $ Z_1 + Z_2> 173 $, the relationship between $\sigma_{ff}^{\gamma\gamma}$ and $\sigma_{bf}^{\gamma\gamma}$ is the opposite (see, e.g., \cite{Shabaev2014}). It was shown in \cite{LM2016} that both the one-photon contribution and the two-photon contribution to the cross section  of $e^+e^-$  pair production are exponentially suppressed at $\beta\ll \beta_0$, where
\begin{equation}
\beta_0=\left[\dfrac{m_e\,(\eta_1+\eta_2)}{2M_p}\right]^{1/3}\,,
\end{equation}
so that the condition $\beta>\beta_0$  must be satisfied. In addition, it was pointed out that the one-photon contribution may become dominant over the two-photon contribution, since it is not suppressed in $\beta$, although it contains the square of the nuclear mass in the denominator \cite{Okun1955}, \S 100 in \cite{BLP}. For $\beta\ll 1$, the multipole expansion is applicable when calculating the one-photon contributions $\sigma_{ff}^{\gamma}$ and $\sigma_{bf}^{\gamma}$ to the pair production cross section. In Refs.~\cite{Lif1935, Okun1955}, the contribution of electric dipole radiation $\sigma_{ff}^{(E1)}$ was found, but this contribution is strongly suppressed by the magnitude of the electric dipole moment of scattered nuclei. Therefore, the contribution of the electric quadrupole radiation may be important.

 Our work is devoted to the calculation of the electric quadrupole contribution  $\sigma_{ff}^{(E2)}$ to $e^+e^-$ pair production cross section and its comparison with other contributions. In addition, we  calculate the contribution of electric dipole radiation and electric quadrupole radiation to the cross sections $\sigma_{bf}^{(E1)}$ and  $\sigma_{bf}^{(E2)}$.

\section{General relations}
For convenience of further discussion, we present a short derivation of the general expression for the one-photon contribution to the $e^+e^-$ pair production  cross section  in collisions of nonrelativistic nuclei. We use a photon propagator in a form convenient for solving nonrelativistic problems,
$$D^{ab}(\omega,\bm k)=-\dfrac{4\pi}{Q^2}(\delta^{ab}-k^ak^b/\omega^2)\,,  \quad D^{0a}=D^{00}=0\,,\quad Q^2= \omega^2-\bm k^2\,.$$
Let $J^\mu$ be the current corresponding to the emission of a virtual photon, $j^\mu$ is the current corresponding to the conversion of a virtual photon into an electron-positron pair. Then the  cross section  $\sigma_{ff}^{\gamma}$ has the form
\begin{align}
&\sigma_{ff}^{\gamma}=\alpha\int\dfrac{d^3p_1\,d^3p_2}{(2\pi)^6}\,\sum_{pol}\, |J^aD^{ab}(\varepsilon_1+\varepsilon_2,  \bm p_1+\bm p_2)\,j^b|^2\,d\varrho_N\,,
\end{align}
where $\bm p_1 $ and $\bm p_2 $ are the momenta of electron and  positron, $\varepsilon_1$  and $\varepsilon_2$ are their energies, integration over $d\varrho_N$  corresponds to integration over the phase space of the final nuclei, and it is assumed that emission of a soft virtual photon does not affect this phase space. We transform the expression for $\sigma_{ff}^{\gamma}$ to the form
\begin{align}
&\sigma_{ff}^{\gamma}=\int\dfrac{d^3k\,d\omega}{(2\pi)^4}\, J^a\,J^{b*}\,D^{ai}(\omega,\bm k) \,D^{bj}(\omega,\bm k) \,P^{ij}(Q)\,\theta(Q^2-4m_e^2)\,d\varrho_N\,,\nonumber\\
&P^{\mu\nu}(Q)=\alpha\,\int\dfrac{d^3p_1\,d^3p_2}{(2\pi)^6}\,(2\pi)^4\,\delta(\varepsilon_1+\varepsilon_2-\omega)\delta(\bm p_1+\bm p_2-\bm k)\,\sum_{pol}j^\mu\,j^{\nu*}\nonumber\\
&=(g^{\mu\nu}-Q^\mu Q^\nu/Q^2)\,{\cal P}(Q)\,,\quad Q^0=\omega\,,\quad \bm Q=\bm k\,,
\end{align}
where $\theta(x)=[1+\mbox{sgn}(x)]/2$ is the Heaviside function. As is known (\S 113 in \cite{BLP}), the function ${\cal P} (Q)$ is
\begin{align}
&{\cal P}(Q)=-\dfrac{\alpha\,}{6\pi}\,\sqrt{1-\dfrac{4m_e^2}{Q^2}}\left(Q^2+2m_e^2\right)\,,\quad Q^2>4m_e^2\,.
\end{align}
Taking the integral over the angles of the vector $\bm k$ and passing from integration over $k$ to integration over $t=k/\omega$, we obtain
\begin{align}
&\sigma_{ff}^{\gamma}=\dfrac{\alpha}{3\pi^2}\,\int\theta(\omega-2m_e)\,\Phi\left(\dfrac{2m_e}{\omega}\right)\,\omega\,d\omega\,\,\bm J\cdot\bm J^{*}\,d\varrho_N\,,\nonumber\\
&\Phi(x)=\int_0^{\sqrt{1-x^2}}dt\,\dfrac{t^2}{1-t^2}\,\sqrt{1-\dfrac{x^2}{1-t^2}}\,\left(2+\dfrac{x^2}{1-t^2}\right)\,\left(1-\dfrac{t^2}{3}\right)\,.
\end{align}
For $x\ll 1$ we have
$$\Phi(x)=-\dfrac{4}{3}\,\ln\,x\,.$$
Since the bremsstrahlung cross    section $d\sigma_{rad}$ reads
\begin{equation}
d\sigma_{rad}=\dfrac{2}{3\pi}\,\omega\,d\omega \,\bm J\cdot\bm J^{*}\,d\varrho_N\,,
\end{equation}
we arrive at the general formula
\begin{align}
&\sigma_{ff}^{\gamma}=\dfrac{\alpha}{2\pi }\,\int\,\theta(\omega-2m_e)\,\Phi\left(\dfrac{2m_e}{\omega}\right)\,d\sigma_{rad} \,.
\end{align}
Note that this formula is valid for any multipolarity of the photon emission. The further problem   reduces to the calculation of $d\sigma_{rad}$, which can be carried out in the framework of classical electrodynamics:
\begin{align}
&d\sigma_{rad}=\dfrac{1}{\omega}
\int {d\,{\cal E}_\omega}\,2\pi\rho\,d\rho\,,
\end{align}
where $d{\cal E}_\omega$  is the spectral intensity of the radiation for a given impact parameter $\rho$. It is convenient to pass from integration over $\rho$  to integration over the parameter $\epsilon$, which characterizes the trajectory of a charged particle in the Coulomb field and is expressed in terms of $\rho $ (see \S 70 in \cite {LL2}) by the relation
\begin{equation}
\epsilon=\sqrt{1+\dfrac{\rho^2}{a^2}}\,,\quad a=\dfrac{\eta_1\eta_2}{\alpha\beta^2\mu}\,,\quad\mu=\dfrac{M_1\,M_2}{M_1+M_2}=M_p\dfrac{A_1\,A_2}{A_1+A_2}\,, 
\end{equation}
where $M_p$  is the proton mass and $A_{1,2}$  are the mass numbers of nuclei. Accounting for  the finite nuclear sizes  reduces to integration over $\epsilon $ in the region
 $\epsilon\geq\epsilon_{min}$, where
$$\epsilon_{min}=\sqrt{1+\dfrac{R^2}{a^2}}\,,\quad R=R_1+R_2\,.$$ 
 
\section{Electric dipole radiation}
The contribution $d\sigma_{rad}^{(E1)}$  of electric dipole radiation to $d\sigma_{rad}$  in a nuclear collision has the form (see \S 70 in \cite{LL2})
\begin{align}
&d\sigma_{rad}^{(E1)}=\dfrac{4\pi^2a^2(\eta_1\eta_2)^2\,{\cal D}_1  }{3\alpha\beta^4\,M_p^2}\,
\omega\,d\omega\,\nonumber\\
&\times e^{-2\pi\nu}\, \int_{\epsilon_{min}}^\infty\left[\big(H_{i\nu}^{(1)\,'}(i\nu\epsilon)\big)^2-\dfrac{(\epsilon^2-1)}{\epsilon^2} \big(H^{(1)}_{i\nu}(i\nu\epsilon)\big)^2\right]\,\epsilon\,d\varepsilon\,,\nonumber\\
&\nu=\dfrac{\omega}{\omega_0}\,,\quad \omega_0=\dfrac{\beta}{a}\,,\quad {\cal D}_1=\left(\dfrac{Z_1}{A_1}-\dfrac{Z_2}{A_2}\right)^2\,.
\end{align}
The leading  logarithmic  contribution to the pair production cross section is determined by the region
 $\epsilon_{min}\ll\epsilon\ll 1/\nu$, in which
$$H^{(1)}_{i\nu}(i\nu\epsilon)\approx \dfrac{2}{i\,\pi}\,\ln\left(\dfrac{1}{\nu\epsilon }\right)\,,\quad H_{i\nu}^{(1)\,'}(i\nu\epsilon)=\dfrac{2}{\pi\nu\epsilon}\,.$$
Using these asymptotics, we take the integral over $\epsilon $ with logarithmic accuracy and obtain
\begin{align}
&d\sigma_{rad}^{(E1)}=\dfrac{16(\eta_1\eta_2)^2\,{\cal D}_1  }{3\alpha\beta^2\,M_p^2}\,\ln\left(\dfrac{\omega_0}{\omega\epsilon_{min}}\right)\,
\dfrac{d\omega}{\omega}\,.
\end{align}
Note that this result is determined by the term $\propto(H_{i\nu}^{(1)\,'})^2$.
The logarithmic contribution $\sigma_{ff}^{(E1)}$  to the cross section of pair production  is determined by the region $m_e\ll \omega\ll \omega_0/\epsilon_{min}$. Taking the integral over $\omega $ with logarithmic accuracy, we get
\begin{align}
\sigma_{ff}^{(E1)}=\dfrac{16\,(\eta_1\eta_2)^2\,{\cal D}_1 }{27\pi\,\beta^2M_p^2}\,\ln^3\left(\dfrac{\omega_0}{m_e\epsilon_{min}}\right) \,.
\end{align}
This formula agrees with the results of \cite{Okun1955} obtained for $R\gg a$  and $R\ll a$  (see also \S 100 in \cite {BLP}, where the case $ R\gg a$ is considered). For $R=a$, the relative velocity $\beta$ equals  $\beta_1$, where
\begin{equation}
\beta_1= \sqrt{\dfrac{(\eta_1+\eta_2)}{2M_p\,R}}>\beta_0\,.
\end{equation}
Note that $\omega_0/\epsilon_{min}\ll 1/R$, since
 $$\dfrac{(R/a)\beta}{\sqrt{1+(R/a)^2}}\ll 1$$ 
for any value of $R/a$. Hence, $\omega\ll 1/R$  and $Q^2\ll 1/R^2$, so there is no need to take into account the form factors of the nuclei  when calculating  $d\sigma_{rad}$.

\section{Electric quadrupole radiation}
The spectral intensity of the electric quadrupole radiation can be calculated in the same way as
electric dipole radiation (see, e.g., \cite{Greiner1976}). The corresponding contribution $d\sigma_{rad}^{(E2)}$ reads
\begin{align}
&d\sigma_{rad}^{(E2)}=\dfrac{ \pi^2(\eta_1\eta_2)^2}{15\,\alpha\,M_p^2 }{\cal D}_2\,e^{-2\pi\,\nu}\,\nu^4\,\dfrac{d\omega}{\omega}\,\int_{\epsilon_{min}}^\infty \dfrac{d\epsilon}{\epsilon}\Bigg\{  \Big[(\epsilon^2-1)\,H_{i\nu}^{(1)\,'}(i\epsilon\nu)+\dfrac{1}{i\nu\epsilon}H^{(1)}_{i\nu}(i\epsilon\nu)\Big]^2\nonumber\\
&+(\epsilon^2-1)^2\Big[H_{i\nu}^{(1)\,'}(i\epsilon\nu)-\dfrac{1}{i\nu\epsilon}H^{(1)}_{i\nu}(i\epsilon\nu)\Big]^2\nonumber\\
&+(\epsilon^2-1)\Big[(\epsilon^2-1)\,H_{i\nu}^{(1)\,'}(i\epsilon\nu)+\dfrac{1}{i\nu\epsilon}H^{(1)}_{i\nu}(i\epsilon\nu)\Big]\Big[H_{i\nu}^{(1)\,'}(i\epsilon\nu)-\dfrac{1}{i\nu\epsilon}H^{(1)}_{i\nu}(i\epsilon\nu)\Big]\nonumber\\   &-3\,\epsilon^2\,(\epsilon^2-1)\,\Big[\dfrac{1}{i\nu\epsilon}\,H_{i\nu}^{(1)\,'}(i\epsilon\nu)+\dfrac{\epsilon^2-1}{\epsilon^2}\,H^{(1)}_{i\nu}(i\epsilon\nu)\Big]^2   \Bigg\}\,,\nonumber\\
&{\cal D}_2=\left[\dfrac{2A_1A_2}{(A_1+A_2)}\left(\dfrac{Z_1}{A_1^2}+\dfrac{Z_2}{A_2^2}\right)\right]^2\,.
\end{align}
Taking the integral with logarithmic accuracy in the region $\epsilon_{min}\ll\varepsilon\ll 1/\nu$, we find
\begin{align}
&d\sigma_{rad}^{(E2)}=\dfrac{4\,(\eta_1\eta_2)^2\,{\cal D}_2}{5\,\alpha M_p^2}\, \,\ln\left(\dfrac{\omega_0}{\omega\epsilon_{min}}\right)\,\dfrac{d\omega}{\omega}\,.
\end{align}
The corresponding contribution of the quadrupole radiation to the cross section $\sigma_{ff}^{\gamma}$ have the form
\begin{align}
&\sigma_{ff}^{(E2)}=\dfrac{4\,(\eta_1\eta_2)^2\,{\cal D}_2}{45\pi M_p^2}\,\ln^3\left(\dfrac{\omega_0}{m_e\epsilon_{min}}\right) \,.
\end{align}
Parametrically, the probability of pair production due  quadrupole radiation of virtual photon is less the probability of pair production due  electric dipole radiation  by the factor $\beta^2$.

\section{The bound-free pair production}
Since the electron-positron pair is produced at distances much less than the Bohr radius, then when calculating the current $j^\mu$  corresponding to the conversion of a virtual photon into an electron-positron pair, we can use the Dirac  spinor $\overline{\Psi}_{1,2}=\psi_{1,2}(0)\,(\phi^+,0)$ as  the electron wave function,  where  $\phi$ is a two-component spinor and $\psi_{1,2}(0)=\sqrt{(m_e\eta_{1,2})^3/(\pi\,n^3)}$ is the nonrelativistic wave function at the origin, $n$ is the principal quantum number. As a result, the cross section of bound-free pair production is
\begin{align}
&\sigma_{bf}^\gamma=\int\dfrac{d^3p}{(2\pi)^3}\, J^a\,J^{b*}\,D^{ai}(\varepsilon +m_e, \bm  p) \,D^{bj}(\varepsilon +m_e,  \bm p) \,P_{bf}^{ij}\,d\varrho_N\,,\nonumber\\
&P_{bf}^{ij}=\alpha\sum_{pol}j^i\,j^{j*}
=\dfrac{\alpha m_e^3\zeta_{3}(\varepsilon+m_e)}{\pi\,\varepsilon}\,(\eta_1^3+ \eta_2^3) \,\delta^{i\,j}\,.
\end{align}
where $\bm p$  and $\varepsilon$  are the momentum and energy of a positron, the Riemann zeta-function $\zeta_{3}=\sum_{n=1}^{\infty}\frac{1}{n^{3}}\approx 1.2$ 
arises from summation over $n$. Integrating over the angles of the vector $\bm p$, we obtain
\begin{align}
&\sigma_{bf}^\gamma=  \alpha\,\zeta_{3}\,(\eta_1^3+ \eta_2^3)\, \int \,\theta(\omega-2m_e)\,  
\dfrac{m_e}{\omega}\,\left(2+\dfrac{4m_e^2}{\omega^2}\right)\sqrt{1-\dfrac{2m_e}{\omega}}\,d\sigma_{rad}\,,
\end{align}
where $\omega=\varepsilon+m_e$. The integral over  $\omega$ converges at $\omega\sim 2m_e$. The contribution of $\sigma^{(E1)}_{bf}$  of electric dipole radiation to the cross section $\sigma_{bf}^{\gamma}$ reads
\begin{align}
&\sigma^{(E1)}_{bf}= \dfrac{416\,\zeta_{3}(\eta_1 \eta_2)^2\,{\cal D}_1}{105\,\beta^2 M_p^2}\,\, (\eta_1^3+ \eta_2^3)\, \ln\left(\dfrac{\omega_0}{m_e\epsilon_{min}}\right) \,.
\end{align}
The contribution $\sigma^{(E2)}_{bf}$  of the quadrupole radiation is
\begin{align}
&\sigma^{(E2)}_{bf}= \dfrac{104\,\zeta_{3}\,(\eta_1\eta_2)^2\,{\cal D}_2 }{175\,M_p^2 }\,(\eta_1^3+ \eta_2^3)\, \ln\left(\dfrac{\omega_0}{m_e\epsilon_{min}}\right) \,.
\end{align}
The following remark should be made here. The characteristic impact parameters, which make the main contribution to the pair production cross section, lie in the interval $a/\nu>\rho>R$, so that 
$$\beta\,\lambda_C\gg \rho\gg R\,.$$
 At distances $r\sim\rho$, the square of the radial part of the Dirac wave function for the bound state has a gain factor
$$\Xi= \left(\dfrac{\lambda_C}{\eta\, r}\right)^{2(1-\sqrt{1-\eta^2})}\approx \left(\dfrac{\lambda_C}{\eta\, r}\right)^{\eta^2} $$
compared to the square of the nonrelativistic radial wave function (see \S 36 in \cite {BLP}).
Therefore, for the applicability of the results obtained above for $\sigma^{(E1)}_{bf}$  and $\sigma^{(E2)}_{bf}$, the condition $\eta^2\ln[\lambda_C/(\eta\,R)]\ll 1$  or $ Z <40 $   must be fulfilled. On the other hand, the condition $Z\gg 1$  must also be fulfilled  so that the capture of an electron does not affect the scattering of nuclei.
 
\section{Discussion of results and conclusion}
As already pointed out, there are two mechanisms of $e^+e^-$ pair production: the two-photon mechanism and the bremsstrahlung (one-photon) mechanism. The contributions of the two-photon mechanism, $\sigma_{ff}^{\gamma\gamma}$ and $\sigma_{bf}^{\gamma\gamma}$, are strongly suppressed by a high degree of $\beta$. The contribution of the one-photon mechanism, $\sigma_{ff}^{(E1)}$ and $\sigma_{bf}^{(E1)}$,  due to   electric dipole radiation of a virtual photon is suppressed by the factor $1/M_p^2$, but enhanced by the factor $1/\beta^2$. However, these contributions contain the factor ${\cal D}_1$  which is very small for all  nuclei   except for the lightest ones, since $Z/A\approx1/2$. For example, when scattering iron on copper ${\cal D}_1=0.8*10^{-4}$, gold on silver ${\cal D}_1=1.2*10^{-3}$, lead on copper ${\cal D}_1=4*10^{-3}$, and gold on lead ${\cal D}_1=3*10^{-5}$.

The contributions $\sigma_{ff}^{(E2)}$ and $\sigma_{bf}^{(E2)}$   of quadrupole radiation to the pair production cross sections do not contain the factor $1/\beta^2 $, however  the factor ${\cal D}_2$ is not small, ${\cal D}_2\approx1$.  Therefore, the electric quadrupole radiation may dominate over the electric dipole radiation and over the contribution of the two-photon mechanism.

\begin{figure}[h]
	\centering
	\includegraphics[width=0.5\textwidth]{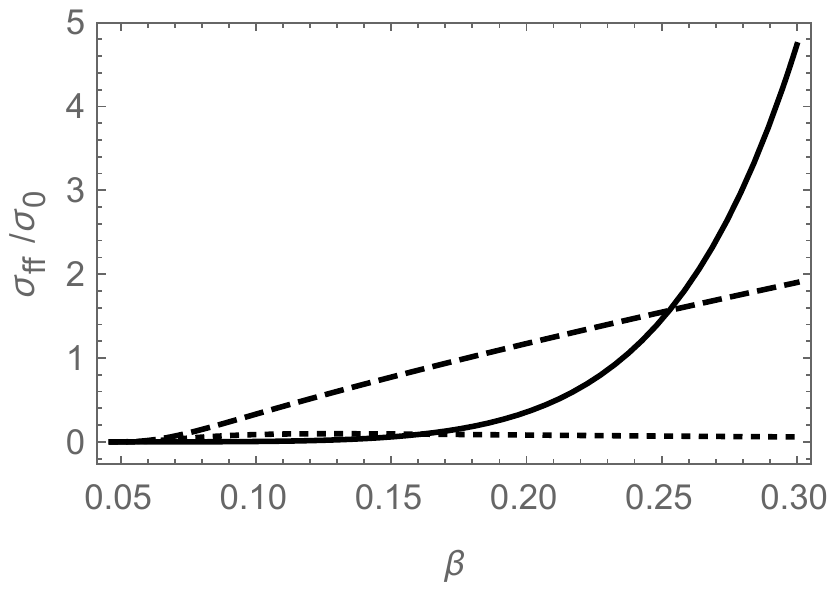}
	\caption{Cross section $\sigma_{ff}$  in units $\sigma_0=(\eta_1\eta_2)^2/(\pi\,M_p^2)$.  Solid curve: the contribution of the two-photon mechanism for $Z_1=Z_2=26$ (iron);  dashed curve: the contribution of quadrupole radiation  for $Z_1=Z_2=26$; dotted curve:  the contribution of electric dipole radiation for $Z_1=47 $ (silver) and $Z_2=26$ (iron).}
	\label{sec123ff}
\end{figure}

\begin{figure}[h]
	\centering
	\includegraphics[width=0.5\textwidth]{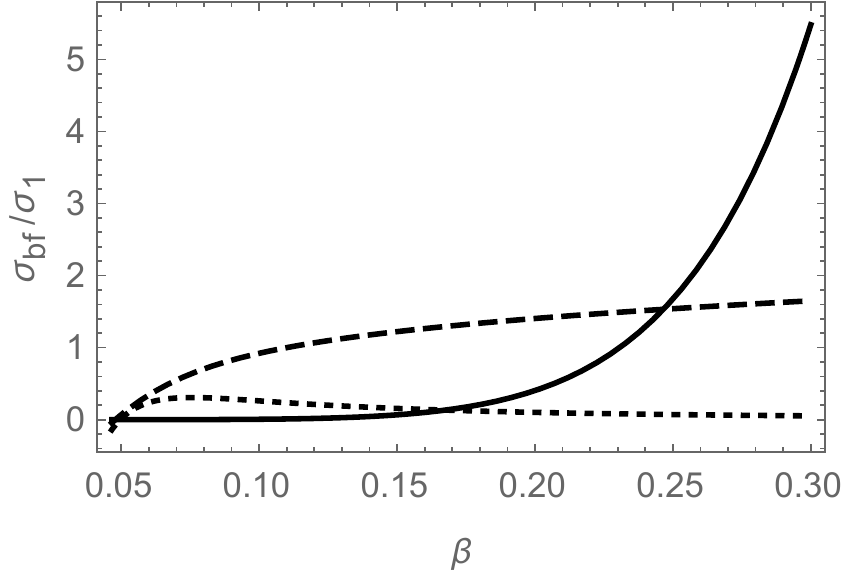}
	\caption{Cross section $\sigma_{bf}$ in units of $\sigma_1=(\eta_1\eta_2)^2(\eta_1^3+\eta_2^3)\zeta_3/M_p^2$. Solid curve: the contribution of the two-photon mechanism for $Z_1=Z_2=26$ (iron);  dashed curve: the contribution of quadrupole radiation  for $Z_1=Z_2=26$; dotted curve:  the contribution of electric dipole radiation for $Z_1=47 $ (silver) and $Z_2=26$ (iron).}
	\label{sec123bf}
\end{figure}

Figure \ref{sec123ff} shows various contributions to the cross section $\sigma_{ff}$  in units of $\sigma_0=(\eta_1\eta_2)^2/(\pi\,M_p^2)$. The solid curve corresponds to the contribution of the two-photon mechanism for $ Z_1 = Z_2 = 26 $ (iron), the dashed curve corresponds to the contribution of quadrupole radiation  for $ Z_1 = Z_2 = 26 $. For comparison, the dotted curve in the same figure shows the contribution of electric dipole radiation to the pair production cross section for $ Z_1 = 47 $ (silver) and $ Z_2 = 26 $ (iron). It is seen that in a wide region of $\beta <1 $ the contribution of the quadrupole radiation is dominant.
Figure \ref{sec123bf} shows the corresponding  contributions to the cross section $\sigma_{bf}$ in units of $\sigma_1=(\eta_1\eta_2)^2(\eta_1^3+\eta_2^3)\zeta_3/M_p^2$ under the same conditions as in Fig.~\ref{sec123ff}. Again, in a wide region of $\beta<1$ the contribution of the quadrupole radiation is dominant.

In conclusion, we have analyzed various contributions to the cross section of $e^+e^-$  pair  production in the collision of nonrelativistic nuclei, taking into account the finite nuclear radius. Both free-free and bound-free cases are discussed. 
It is shown that  the interaction between nuclei  in the collision process is of fundamental importance, since the bremsstrahlung (one-photon) contribution to the cross section can dominate due to the quadrupole emission of a virtual photon.


\begin{thebibliography}{99}
	
	\bibitem{LandLif1934}
	L.~D. Landau, E.~M. Lifshitz, Physik. Z. Sowjetunion. {\bf 6}, 244 (1934) .
	
	\bibitem{Racah1936}
	G.~Racah, Il Nuovo Cimento (1924-1942) {\bf 13~(2)}, 66 (1936).
	
	\bibitem{BaltMcL1998}
	A.~J. Baltz, L.~D. McLerran, Phys.~Rev.~C {\bf58},1679  (1998).
	
	\bibitem{IvaScSe1999}
	D.~Y. Ivanov, A.~Schiller, V.~G. Serbo, Phys. Lett. {\bf B 454}, 155 (1999).
	
	\bibitem{EiReScG1999}
	U.~Eichmann, J.~Reinhardt, S.~Schramm, W.~Greiner, Phys. Rev. A {\bf 59~(2)}, 1223 (1999).
	
	\bibitem{SegeWel1999}
	B.~Segev, J.~C. Wells, Phys.~Rev.~C {\bf 59},  2753 (1999).
	
	\bibitem{Lee2000}
	R.~N. Lee, A.~I. Milstein, Phys. Rev. A {\bf61}, 032103  (2000).
	
	\bibitem{BauHeTr2007}
	G.~Baur, K.~Hencken, D.~Trautmann, Phys. Rept. {\bf 453}, 1 (2007).
	
	\bibitem{UFNKM2019} P.A. Krachkov, A.I. Milstein,
		Usp. Fiz. Nauk {\bf 189}, 359 (2019) [Phys. Uspekhi. {\bf 62 (4)}, 340 (2019)].
		
	\bibitem{Lif1935}E.M. Lifshitz, Physik. Z. Sowjetunion {\bf 7}, 385 (1935).
	
	\bibitem{Okun1955} L. B. Okun, Dokl. Akad. Nauk SSSR {\bf89}, 833 (1953).
	
\bibitem{BLP}
V.B.~Berestetskii, E.M.~Lifshitz, L.P.~Pitaevskii, {\it Quantum Electrodynamics}, Course of theoretical physics (Butterworth-Heineman, 1982).
	
		
\bibitem{Greiner1976}	J. Reinhardt, G.~Soff and W. Greiner, Z. Phys. {\bf A276}, 285 (1976).
		
	\bibitem{Greiner1985}
	W.~Greiner, B.~M{\"u}ller, J.~Rafelski, {\it Quantum Electrodynamics of Strong
	Fields} ( Berlin: Springer, 1985).
	
\bibitem{shabaev2020}	R.~V. Popov, V.~M. Shabaev, D. A. Telnov, Tupitsyn {\it et al.}, Phys. Rev. D {\bf 102}, 076005 (2020).
	
	\bibitem{LeeMingulov2016}
	R.~N. Lee, K.~T. Mingulov, Phys.
	Lett. B {\bf 757}, 207 (2016).
	
	
	\bibitem{LM2016} R.N. Lee   and A. I. Milstein,
		Physics Letters B {\bf 761},  340 (2016).
	
	\bibitem{Shabaev2014}
	I.~A. Maltsev, V.~M. Shabaev, I.~I. Tupitsyn, A.~I. Bondarev, Y.~S. Kozhedub,
	G.~Plunien, T.~St\"ohlker, Phys. Rev. A {\bf91}, 032708 (2015).
	
	\bibitem{LL2} L.D.~Landau,  E.M.~Lifshitz, {\it The Classical Theory of Fields}, Course of theoretical physics (Butterworth-Heineman, 1982).
	
	
	
\end{thebibliography}
\end{document}